\documentclass{article}

\usepackage{arxiv}

\usepackage[utf8]{inputenc} 
\usepackage[T1]{fontenc}    
\usepackage{hyperref}       
\usepackage{url}            
\usepackage{booktabs}       
\usepackage{nicefrac}       
\usepackage{microtype}      
\usepackage{lipsum}
\usepackage{graphicx}
\usepackage{cite}
\usepackage{amsmath,amssymb,amsfonts}
\usepackage{algorithmic}
\usepackage{graphicx}
\usepackage{textcomp}
\usepackage{url}
\usepackage{multirow}
\usepackage{subcaption} 
\usepackage{xcolor}

\title{Uncovering Neuroimaging Biomarkers of Brain Tumor Surgery with AI-Driven Methods}

\author{
 Carmen Jiménez-Mesa \\
  Department of Communication Engineering\\
  University of Málaga\\
  Spain \\
   \And
 Yizhou Wan \\
  Department of Clinical Neurosciences\\
  University of Cambridge\\
  United Kingdom \\
  \And
 Guilio Sansone \\
  Department of Neuroscience\\
  University of Padova\\
  Italy \\
    \And
 Francisco J. Martinez-Murcia \\
  Department of Signal Theory, Telematics \\
  and Communications, University of Granada\\
  Spain \\
    \And
 Javier Ramirez \\
  Department of Signal Theory, Telematics \\
  and Communications, University of Granada\\
  Spain \\
    \And
 Pietro Lio \\
  Department of Computer Science and Technology\\
  University of Cambridge\\
  United Kingdom \\
    \And
 Juan M. Gorriz \\
  Department of Signal Theory, Telematics \\
  and Communications, University of Granada\\
  Spain \\
    \And
 Stephen J. Price \\
  Department of Clinical Neurosciences\\
  University of Cambridge\\
  United Kingdom \\
    \And
 John Suckling \\
  Department of Psychiatry\\
  University of Cambridge\\
 Cambridge and Peterborough NHS Foundation Trust \\
  United Kingdom \\
    \And
 Michail Mamalakis \\
  Department of Psychiatry\\
  Department of Computer Science and Technology \\
  University of Cambridge\\
  United Kingdom \\
  \texttt{mm2703@cam.ac.uk} \\  
}

\begin{document}
\maketitle
\begin{abstract}
Brain tumor resection is a highly complex procedure with profound implications for survival and quality of life. Predicting patient outcomes is crucial to guide clinicians in balancing oncological control with preservation of neurological function. However, building reliable prediction models is severely limited by the rarity of curated datasets that include both pre- and post-surgery imaging, given the clinical, logistical and ethical challenges of collecting such data. In this study, we develop a novel framework that integrates explainable artificial intelligence (XAI) with neuroimaging-based feature engineering for survival assessment in brain tumor patients. We curated structural MRI data from 49 patients scanned pre- and post-surgery, providing a rare resource for identifying survival-related biomarkers. A key methodological contribution is the development of a global explanation optimizer, which refines survival-related feature attribution in deep learning models, thereby improving both the interpretability and reliability of predictions.
From a clinical perspective, our findings provide important evidence that survival after oncological surgery is influenced by alterations in regions related to cognitive and sensory functions. These results highlight the importance of preserving areas involved in decision-making and emotional regulation to improve long-term outcomes. From a technical perspective, the proposed optimizer advances beyond state-of-the-art XAI methods by enhancing both the fidelity and comprehensibility of model explanations, thus reinforcing trust in the recognition patterns driving survival prediction.
This work demonstrates the utility of XAI-driven neuroimaging analysis in identifying survival-related variability and underscores its potential to inform precision medicine strategies in brain tumor treatment.
\end{abstract}


\section{Introduction}\label{sec:introduction}
Gliomas, the most frequent primary brain tumors, vary in aggressiveness, prognosis, and histopathology. Their treatment often involves surgical resection, followed by radiotherapy and chemotherapy. The extent of resection significantly affects survival, with surgery needing to balance tumor removal and brain function preservation \cite{Dadario2021}, a principle often referred to as onco-functional balance. Beyond the immediate surgical outcome, post-operative brain reorganisation plays a central role in functional recovery. However, the mechanisms underlying these structural and functional adaptations remain insufficiently understood \cite{drewes2018perioperative}. A more accurate characterization of these processes is essential for guiding clinical decisions, improving rehabilitation, and ultimately enhancing patient survival and quality of life.

Structural Magnetic Resonance Imaging (sMRI) provides high-resolution insights into the effects of tumor resection on brain structure, but its high dimensionality and complexity pose major analytical challenges. Machine learning (ML) techniques, particularly dimensionality reduction methods such as Principal Component Analysis (PCA) \cite{greenacre2022principal} or  Uniform Manifold Approximation and Projection (UMAP) \cite{mcinnes2018umap}, allow for the extraction of low-dimensional representations that capture meaningful structural variations. These representations facilitate the identification of hidden patterns that may be otherwise invisible in conventional analyses. At the same time, eXplainable Artificial Intelligence (XAI) frameworks, such as feature attribution methods and model interpretability frameworks  \cite{gorriz2023computational,mamalakis}, are increasingly recognised as essential to translate ML findings into clinically interpretable biomarkers, enabling trust and adoption in medical practice.

Despite these advances, most existing studies in brain tumour research have focused on pre-operative imaging, diagnosis, or histological classification. Much less attention has been given to post-surgical structural changes and their relationship with survival, in part due to the scarcity of longitudinal datasets covering both pre- and post-operative stages. The dataset collected and used in this study provides a rare opportunity to directly investigate these dynamics, offering insights into how surgery reshapes brain structure and how such changes relate to long-term outcomes.


In this work, we introduce a novel computational framework that combines neuroimaging-based feature engineering with a global explanation optimizer to investigate structural brain reorganization in glioma patients. The framework is designed to identify survival-related biomarkers while enhancing the stability, fidelity, and clarity of model explanations, thereby minimizing inter-method variability. Utilizing a uniquely curated dataset of pre- and post-surgery sMRI scans, we further examine how surgery-affected brain regions influence survival outcomes. Our ultimate goal is to provide clinically actionable insights that can guide surgical decision-making, refine risk stratification, and support personalized rehabilitation strategies.

The key contributions of this study are:
\begin{itemize}
\item A global explanation optimizer that strengthens the reliability, fidelity, and clarity of survival-related neuroimaging biomarkers.
\item An integrated framework that combines latent-space feature engineering with XAI to provide interpretable assessments of post-surgical brain reorganization
\item Clinical insights into survival and recovery, delivering actionable guidance to neurosurgeons for optimizing surgical strategies, minimizing complications, and tailoring patient-specific post-operative care.
\end{itemize}


To the best of our knowledge, this is the first study to systematically integrate latent-space analysis of sMRI with an XAI optimization framework in the context of brain tumor surgery. By addressing both methodological and clinical challenges, our work goes beyond diagnosis to model how surgery impacts brain structure and survival. This positions our approach at the intersection of methodological innovation and clinical applicability, with direct implications for improving both acute surgical outcomes and long-term patient management.

\section{Related work} 
Machine learning techniques have shown promising results in brain tumor analysis and outcome prediction for neurosurgical patients. Compared to conventional statistical methods, ML algorithms have demonstrated superior performance in predicting postoperative complications \cite{van2019machine,neumann2024routine} and inpatient length of stay \cite{muhlestein2019predicting}.  Beyond clinical outcomes, ML-based approaches have also been widely applied to neuroimaging tasks such as brain tumor segmentation and classification \cite{pugalenthi2019evaluation,diaz2021deep}, often achieving state-of-the-art accuracy. More recent studies have started to explore pre- and post-operative MRI data, for instance to predict recovery trajectories or assess surgical effects on brain anatomy \cite{xavier2021pre,nalepa2023deep,li2025machine}. Similarly, ML models have been employed to predict long-term neurosurgical outcomes, including survival, recurrence, and symptom progression \cite{senders2018machine,abualnaja2024machine}. These works highlight the increasing role of ML in neurosurgery, though the majority remain focused on diagnostic, segmentation, or histopathological classification tasks, rather than on structural reorganization after surgery.

Latent space representations, such as those derived from PCA or other manifold learning methods, have been shown to capture complex patterns in neuroimaging data that are not easily observable in raw high-dimensional spaces. They have been successfully applied in tasks ranging from correlation representation learning in multi-modal MRI segmentation \cite{akbari2016imaging,zhou2021latent} to dimensionality reduction for group-level analyses. Such methods enable interpretable visualization and clustering of subtle neuroanatomical variations. However, to the best of our knowledge, no previous studies have leveraged latent spaces to systematically investigate longitudinal structural changes in brain tumors before and after surgery, nor their relationship with survival. This gap motivates the present study.

Alongside dimensionality reduction, the development of XAI methods has been pivotal in translating ML findings into clinically actionable knowledge \cite{mine2,surv1,mamalakis, akgundougdu2025explainable}. Local XAI methods provide interpretations of individual model predictions, whereas global methods offer cohort-level insights into the model's overall decision-making process, thereby enhancing our understanding of its behavior across populations. In neuroimaging, XAI has been used to highlight relevant regions or modalities associated with tumour detection, disease progression, and prognosis, thereby increasing transparency in clinical AI. Nonetheless, a persistent challenge is the variability of explanations across methods: different local or global XAI techniques often yield inconsistent attributions, which may reduce trust in model-derived biomarkers and hinder clinical translation \cite{mamalakis}. To the best of the authors' knowledge, no global optimal solution exists to address inter-method variability in explanations. Bridging this gap is a key objective of our proposed work.

In summary, while ML has been widely applied to neurosurgery and brain tumor analysis, latent space methods remain underexplored for modelling structural reorganization, and current XAI techniques lack robustness when applied to survival-related neuroimaging biomarkers. Our work bridges these gaps by (i) introducing a latent-PCA framework to capture post-surgical structural changes in glioma patients, and (ii) proposing a global explanation optimizer to mitigate inter-method variability in XAI, thereby offering more stable and clinically interpretable biomarkers of survival.

\section{Methods} 
\begin{figure*}
\centering
\includegraphics[width=1.\textwidth]{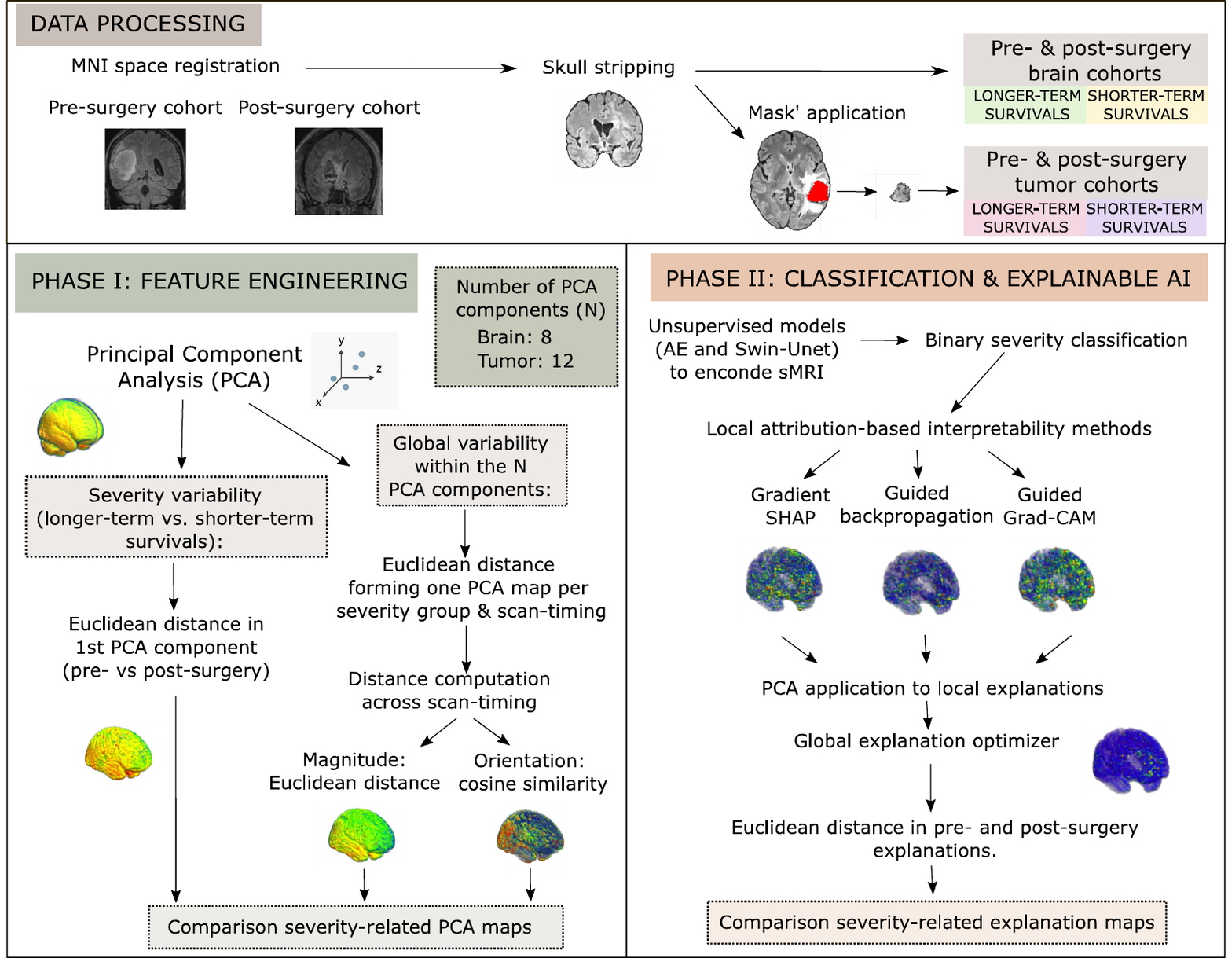}
\caption{Framework proposed. Imaging processing is followed by two different analysis. Phase I: in the feature engineering study PCA components are extracted from the different cohorts analyzing variability between and within groups. Phase II: to enhance interpretability and robutness of the outcomes, an analysis is conducted by means of a binary classifier where three different XAI techniques are applied: Gradient-SHAP, Guided-Backpropagation and Guided-GradCAM. Their outputs serve as input to a global explanation optimizer, generating a map of the most relevant global patterns for each severe condition.} \label{fig2}
\end{figure*}

This work combines latent space feature engineering and XAI methods to identify biomarkers related to surgical outcomes. A summary of the implemented framework is presented in Fig.~\ref{fig2}.  The main part consists of two phases: feature engineering through dimensionality reduction and a global explanation optimizer integrated with DL networks and XAI methods.

\subsection{Phase I: Feature engineering based on dimensionality reduction}

We utilized PCA to extract the most relevant patterns of variation across the brain and tumor cohorts, considering four groups defined by time (pre- vs. post-surgery) and survival (longer-term vs. shorter-term). Two main approaches were used: first, analyzing PCA component variability across groups, and second, quantifying variability across PCA components (see Fig.~\ref{fig2}):

\subsubsection{First PCA component variability across groups} The first PCA component, representing the highest variance, was compared across groups (shorter-term and longer-term survivals) to identify dominant differentiation patterns. To do so, spatial variability between the two conditions (pre- and post-surgery) was compute using voxel-wise Euclidean distance. The Euclidean distance between two PCA-transformed representations, $\mathbf{p}_{A}$ and $\mathbf{p}_{B}$, can be mathematically described as:
\begin{equation}
d_E = \|\mathbf{p}_{A} - \mathbf{p}_{B}\|_2 = \sqrt{\sum_{i=1}^{K} (p_{A, i} - p_{B, i})^2}
\end{equation}
where a larger distance indicates greater structural change.
The comparison of these variability maps allow to assess whether the PCA has captured meaningful group distinctions.  

\subsubsection{Variability quantification across PCA Components} Local variability maps were generated by computing voxel-wise Euclidean distances across the $k$ PCA components of each of the four subgroups, summarizing the total magnitude of variations captured by PCA and quantifying regional brain variability. This approach enabled the estimation of global variability within groups (pre- vs. post surgery) by analyzing both magnitude (Euclidean distance) and orientation (cosine similarity) from the local maps. The cosine similarity can be mathematically described as:
\begin{equation}
S_{\cos} = \frac{\mathbf{p}_{A} \cdot \mathbf{p}_{B}}{\|\mathbf{p}_{A}\| \|\mathbf{p}_{B}\|}
\end{equation}

Once this is done, the global maps of shorter-term and longer-term survivals can be compared to assess spatial variability.

\subsection{Phase II: Feature identification based on cohort-level explanations, integrated with DL networks and tailored to survival classification}

Fig.~\ref{fig2} illustrates the explainable AI framework developed to identify global (cohort-level) patterns associated with survival outcomes following brain tumor surgery. Given the limited size of our clinical dataset and the need to avoid overfitting, we first trained a generalized unsupervised model on a large, heterogeneous dataset of structural MRI brain tumor scans. This encoder–decoder architecture learned the distribution of sMRI data, capturing variability and heterogeneity across patients to reduce bias in downstream analyses.
Building on this foundation, a binary classification model was trained and validated to distinguish patients with shorter versus longer survival. To optimize performance, we systematically evaluated different strategies, including freezing versus fine-tuning encoder layers and conducting an ablation study of alternative network architectures derived from the unsupervised stage. Cohort-level explanations were then integrated into the survival classification task using our proposed global explanation optimizer, which enhanced both the clarity and the consistency of global survival-related patterns.

\subsubsection{Unsupervised learning of structural MRI} \label{sec:xai_1} Two deep learning architectures were employed in the unsupervised learning stage: a convolutional autoencoder (AE) with three encoder and three decoder blocks, and the Swin-Unet \cite{rec}. Both models were trained to reconstruct full pre- and post-operative 3D structural MRI scans in an unsupervised setting. To assess reconstruction performance and generalization capacity, we performed an ablation study comparing two different cohort training strategies. Further implementation details are provided in Section~\ref{sec:implementation}.

\subsubsection{Survival classification of structural MRI}\label{sec:binary} For the survival classification task, we used the encoder components of the previously trained unsupervised AE and Swin-Unet models. An ablation study was conducted to evaluate different output layer configurations: (i) a three-layer multilayer perceptron (MLP) for binary classification, and (ii) a cross-attention (Attention) mechanism applied to the four encoder stages of the Swin-Unet \cite{rec}.
We explored three training strategies: (1) freezing the encoder (freeze) and training only the output layer, (2) fine-tuning the encoder by unfreezing its weights (unfreeze), and (3) re-initializing and jointly training both the encoder and the output layer (full training).

\subsubsection{Global explanations of structural MRI} To enhance interpretability in the survival classification task, we used six local attribution-based methods: Guided Backpropagation \cite{grad}, Guided GradCam \cite{cam}, and Gradient Shap \cite{gradshap}, Input × Gradient \cite{1}, Integrated Gradients\cite{1}, and Kernel SHAP \cite{gradshap}. The goal was to uncover global patterns distinguishing between longer-term and shorter-term survivals outcomes by generating global explanations from pre- and post-surgery sMRI. 
To achieve this, we first estimated the global (cohort-level) pre-surgery and post-surgery explanations using the six different local explanation methods. We then applied PCA to the local explanations generated by each of these XAI methods to obtain a globalized representation. Finally, Euclidean distances were used to quantify differences between the global pre- and post-surgery explanations.
To assess the accuracy of these explanations, we evaluated sparseness \cite{sparse} and faithfulness \cite{fid}. These explainability metrics were computed using the software developed by \cite{quan}, a comprehensive toolkit designed to collect, organize, and assess various performance metrics proposed for XAI methods. We note that a zero baseline (``black'') and 20 random perturbations were used to compute the faithfulness score.

\subsubsection{The proposed global explanation optimizer of structural MRI} 
To identify potential biomarkers, reduce inter-method global explanation variability, and extract actionable insights for improving surgical outcomes, we aimed to generate a global explanation for the binary survival classification task. To this end, we proposed a global explanation optimizer, building on the methodology introduced by \cite{mamalakis} for optimizing explanation representations.
Our framework follows the foundational design of the original approach, including a non-linear encoder-decoder architecture (Swin-Unet) and a multi-objective cost function. A key distinction in our implementation lies in the evaluation strategy: we assess the optimized global explanation by comparing it to the first principal component extracted via PCA on the sMRI data. This comparison enables quantitative assessment of structural relevance using the Structural Similarity Index Measure (SSIM).

We extracted the first three principal components via PCA from the total cohort saliency maps, generated using three of the six widely used attribution methods employed in this study: Guided Backpropagation, Guided Grad-CAM, and Gradient SHAP. These components, along with their weighted average, calculated according to the procedure described in \cite{mamalakis}, were used as four inputs to the proposed global explanation optimizer.

The cost function guiding the optimization integrates three key components: sparseness, as defined in \cite{sparse}; faithfulness \cite{fid}, to ensure consistency with model predictions; and similarity, to align the optimized explanation with a structural representation. This composite objective supports the generation of explanations that are both interpretable and clinically meaningful.

The resulting SSIM score between the optimized global explanation and the first PCA component of the structural MRI inputs is reported as follows:

\begin{equation}
loss_{sim}(\boldsymbol{x},\boldsymbol{y})= \frac{(2 \mu_x \mu_y + c_1)(2 \sigma_{xy}+c_2)}{(\mu_x^2+\mu_y^2+c_1)(\sigma_x^2+\sigma_y^2+c_2)}
\label{1}
\end{equation}
where $\boldsymbol{x}$ represents the derived explanation by the global optimizer, $\boldsymbol{y}$ denotes the first component of PCA of the structural MRI, $\mu_{x}$ indicates the average of $\boldsymbol{x}$, $\sigma_x^2$ signifies the variance of $\boldsymbol{x}$, $\sigma_{xy}$ represents the covariance of $\boldsymbol{x}$ and $\boldsymbol{y}$, and $c_1$ and $c_2$ are two parameters utilized to stabilize the division with a weak denominator \cite{ssim}.  
The total loss function was given by:
\begin{align}
    loss_{total}(\boldsymbol{x},\boldsymbol{y}) = & \, l_{1} \frac{1}{M_{faith}(f,g;\boldsymbol{x})} 
    + l_{2} M_{sparse}(f,g;\boldsymbol{x}) \notag \\ 
    & + l_3 loss_{sim}(\boldsymbol{x},\boldsymbol{y}) 
    \label{3}
\end{align}
where $M_{\text{sparse}}$, $M_{\text{faith}}$ are the metrics for sparseness \cite{sparse} and faithfulness \cite{fid}, respectively and the $g$ global explanation for the network $f$.

\subsection{Summary of the proposed framework}
The proposed framework provides a unified pipeline to analyze post-surgical brain structural changes and identify survival-related biomarkers from MRI data. As shown in Fig.~\ref{fig2}, the workflow has a clear flow from raw imaging data to clinically interpretable outcomes. Before the main analysis, all sMRI images are preprocessed to ensure consistency across subjects. This includes spatial alignment, skull stripping, and masking of tumor regions. These steps harmonize the images and reduce variability unrelated to brain structure. The framework then proceeds in two complementary phases, combining latent-space feature engineering and explainable AI methods to extract meaningful patterns and biomarkers:
\begin{enumerate}
    \item Phase I – Latent-Space Feature Engineering: Pre- and post-surgery MRI scans are transformed into low-dimensional latent spaces using PCA. This step captures the most relevant patterns of variability across the brain and tumor cohorts, grouped by time (pre- vs. post-surgery) and survival (longer-term vs. shorter-term). By quantifying both local and global variability across PCA components and groups, this phase provides a comprehensive view of structural changes induced by surgery.
    \item Phase II – Cohort-Level Feature Identification via XAI: Latent representations are then used to train survival classifiers based on DL encoders. Multiple local XAI techniques are applied to the trained models to produce individual-level explanations. These explanations are combined and optimized using our global explanation optimizer, yielding stable and interpretable cohort-level maps of brain regions associated with survival outcomes.
\end{enumerate}

The framework therefore bridges unsupervised feature extraction and explainable deep learning, enabling the identification of meaningful structural patterns while ensuring robustness and interpretability. The final outputs are global explanation maps, highlighting key brain regions and tumor areas linked to survival.

\section{Experimental settings}
\subsection{Dataset}

The main dataset was from Addenbrooke's Hospital (Cambridge, UK) which consists of 49 MRI T2-weighted scans acquired both before and after surgical resection of the tumour. These scans were spatially normalized to MNI space using SPM12 (\url{fil.ion.ucl.ac.uk/spm/}) and resampled to a 1$\times$1$\times$1 mm$^3$ resolution resulting in final image dimensions of 157$\times$189$\times$156 mm. Skull-stripping was performed using HD-BET \cite{isensee2019hdbet}. Patients were categorized into two outcome groups: longer-term (32) and shorter-term (17) survivals. Most patients (42, $85\%$) had a glioblastoma, but there were also cases of astrocytoma (1), gliosarcoma (3) and others (3). The shorter-term survival group comprised patients who had died within 10 months after the postoperative scan. In contrast, the longer-term group included those who survived for more than 10 months. All individuals gave written informed consent to participate, and the use of their data for clinical research was approved by the Research Ethics Committee (REC reference: 19/WM/0152).

In Phase II, an additional dataset from the 2025 Brain Tumor Segmentation (BraTS) Glioma Challenge \cite{brats} was employed. Hereafter, we refer to this dataset as BraTS2025. This dataset comprises pre- and post-treatment T2-weighted MRI scans. We used a total of 1453 images (1251 pre-treatment and 202 post-treatment). These scans were used to train the unsupervised learning models (see \ref{sec:xai_1}). Demographic information was not provided for this dataset.

\subsection{Implementation details}\label{sec:implementation}
For PCA computation, sMRI scans were vectorized and standardized with zero-mean, unit-variance scaling. No intensity normalisation was applied to the tumour masks due to their binary nature and spatial variability. PCA outcomes were normalized using min-max scaling to $[0,1]$. Eight components were selected for brain images and 12 for tumor images based on cumulative variance with $8$ components explaining over $80\%$ of variance in both severity conditions. Tumor images required $12$ components for similar variance. 

To enhance reproducibility and facilitate result interpretation, the outcomes of these and subsequent analyses were mapped onto the Human Connectome Project (HCP) HCP-MMP1 atlas \cite{glasser2016multi}.

For the unsupervised learning task, a fixed-step learning rate ($5 \times 10^{-4}$) and the Adam optimizer \cite{adam} were used to minimize a SSIM-based loss function \cite{ssim} (see \ref{1}). The learning rate remained constant, with early stopping after 10 epochs of no improvement (max 200 epochs). Two cohort training strategies were evaluated: (i) using only the Addenbrooke’s Hospital dataset, and (ii) combining the Addenbrooke’s Hospital and BraTS2025 datasets. For the Addenbrooke’s Hospital dataset, the 96 available scans were randomly shuffled and divided into five folds for cross-validation (CV) across the entire cohort. In the combined dataset scenario, a 60/40 training/validation (1453 and 96 3D-MRI scans) split was employed.

\begin{figure}
\centering
\begin{subfigure}[t]{\columnwidth}
    \centering
\includegraphics[width=0.95\columnwidth]{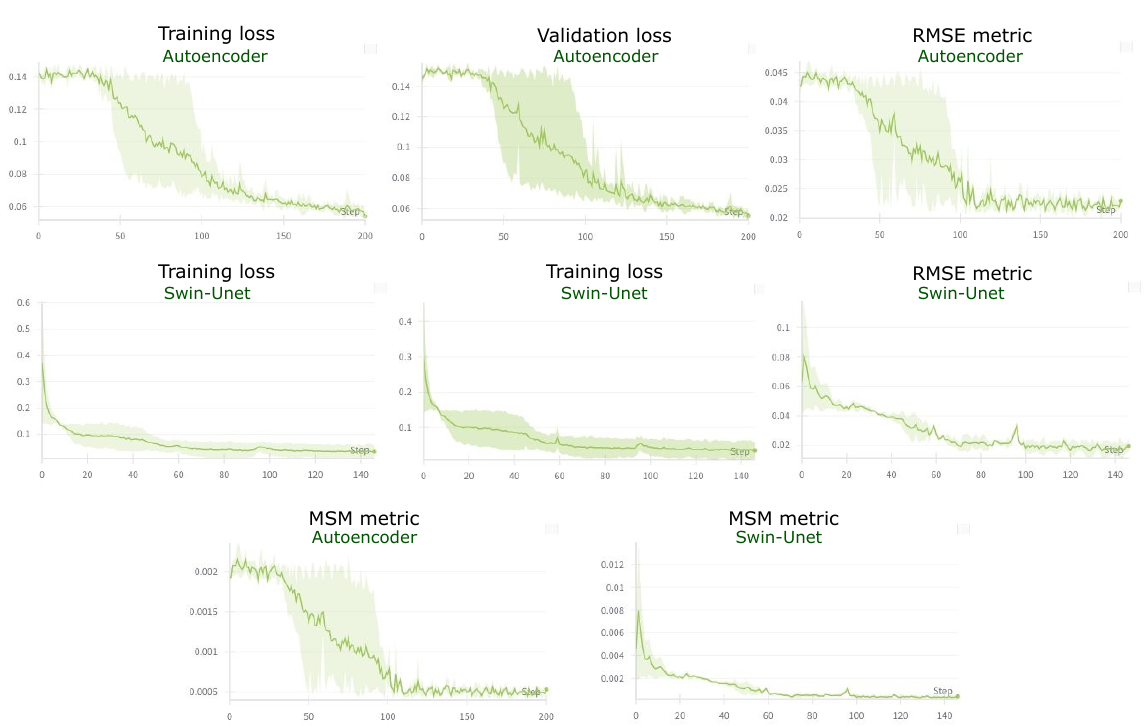}
    \caption{Hyperparameter learning rate tuning of unsupervised learning architectures.}
    \label{fig:unsupervised}
\end{subfigure}
\begin{subfigure}[t]{\columnwidth}
    \centering
    \includegraphics[width=0.95\columnwidth]{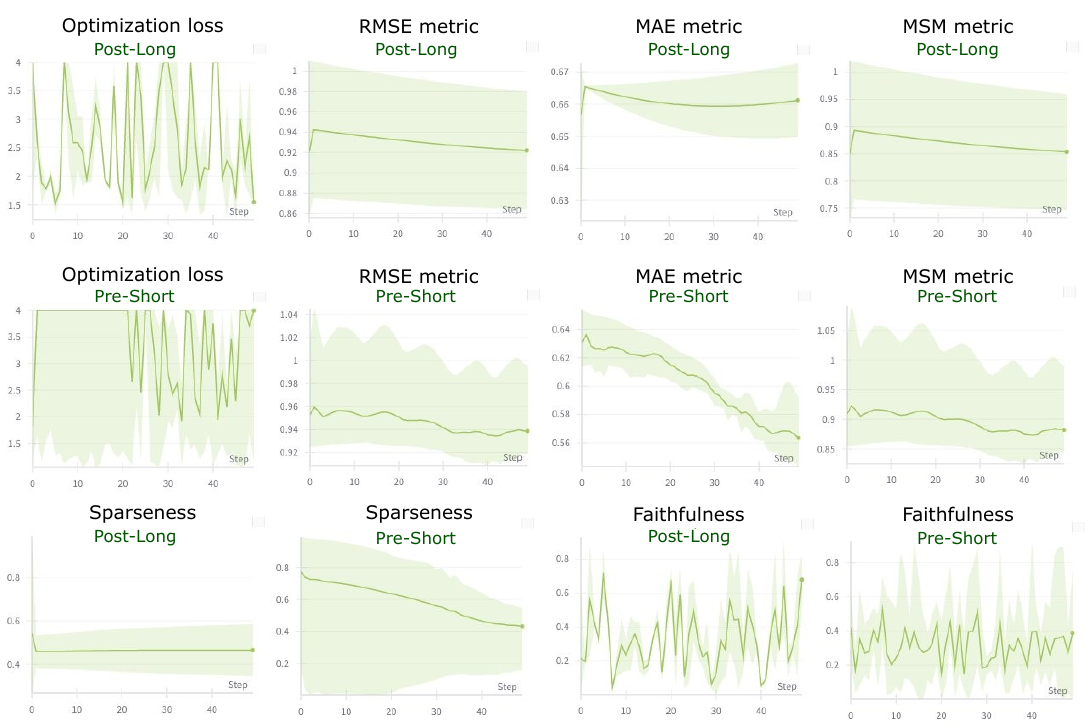}
    \caption{Hyperparameter tuning of various combinations of cost functions ($l_1$, $l_2$, and $l_3$) for global explanation models in both post-surgery longer-term and pre-surgery shorter-term survival groups.}
    \label{fig:xai}
\end{subfigure}
\caption{Examples of hyperparameter tuning results for (a) unsupervised learning, (b) global explanation models on structural MRI. Abbreviations: RMSE–Root Mean Squared Error, MSM–Mean Squared Magnitude, MAE–Mean Absolute Error.}
\label{fig31}
\end{figure}

For survival classification, sparse categorical cross-entropy was used as the loss function, optimized with Adam. The learning rate was constant for the first 100 epochs and then reduced by a factor of 0.1 every 100 epochs. Early stopping was applied after 100 epochs of no improvement (max 400 epochs). A 5-fold CV was used.
Both tasks employed data augmentation, including rotation ($[-15^\circ, 15^\circ]$), width/height shift (up to 20 pixels), and intensity shift (up to 20\%). Hyperparameter tuning tested learning rates: $5 \times 10^{-2}$, $5 \times 10^{-3}$, $5 \times 10^{-4}$, and $5 \times 10^{-5}$ (see Fig. \ref{fig31}a.). The XAI task used the Adam optimizer, but no data augmentation. The cost function was \eqref{3}. For 3D tasks, training lasted up to 100 epochs, with early stopping after 10 epochs of no improvement beyond the first 50. Hyperparameter tuning tested the same learning rates as previously and various combinations of the $l_1$, $l_2$, and $l_3$ parameters in \eqref{3} with the best combination of parameters determined as $l_1 = 0.4$, $l_2 = 0.3$, $l_3 = 0.3$ and a learning rate $5 \times 10^{-5}$ (see Fig. \ref{fig31}b.). 
Codes were implemented in Python using PyTorch and trained on one A100 GPU with 64~GB RAM. It will be publicly available on GitHub.

\subsection{Explanation Quality Metrics}
A critical component of this study is the evaluation of \textit{how accurate and comprehensive an explanation is}. To this end, we focus on two essential metrics: faithfulness and complexity. One intuitive and widely adopted approach for assessing explanation quality is to examine how well it captures the behavior of a predictive model under input perturbations~\cite{f}.

\subsubsection{Faithfulness Metric} Let \( f \) denote a deep neural network, and let \( \boldsymbol{x} \in \mathbb{R}^d \) represent an input with \( d \) features. We aim to assess whether the attribution scores—also known as feature importance scores—accurately reflect the impact of each feature on the model's output.

Consider a subset \( S \subseteq \{1, 2, ..., d\} \) of input features, and let \( \boldsymbol{x}_S \) denote the corresponding sub-vector of \( \boldsymbol{x} \), with \( \boldsymbol{x}_S^f \) being the baseline (reference) values for those features. If \( g(f, \boldsymbol{x}) \in \mathbb{R}^d \) is the attribution vector provided by explanation method \( g \), then the faithfulness is measured by the Pearson correlation between the sum of attributions for the features in \( S \) and the change in the model's output when those features are set to baseline:
\begin{equation}
M_{\text{faith}}(f, g; \boldsymbol{x}) = \text{corr}_S\!\left( \sum_{i \in S} g(f, \boldsymbol{x})_i,\; f(\boldsymbol{x}) - f(\boldsymbol{x}[\boldsymbol{x}_S = \boldsymbol{x}_S^f]) \right)
\label{eq:faithfulness}
\end{equation}
where \( \boldsymbol{x}_F = \boldsymbol{x} \setminus \boldsymbol{x}_S \) denotes the unchanged features.

\subsubsection{Sparseness Metric} To quantify the complexity of an explanation, we evaluate the sparseness of the attribution vector. Sparseness indicates whether the explanation highlights only the most relevant features, which is desirable for interpretability.

We use the \textit{Gini Index}, a well-established measure of inequality, to assess sparseness~\cite{spar}. Given a non-negative vector \( \boldsymbol{v} \in \mathbb{R}^d_{\geq 0} \), let \( v_{(k)} \) be the \( k \)-th smallest value after sorting. The Gini Index is defined as:

\begin{equation}
G(\boldsymbol{v}) = 1 - 2 \sum_{k=1}^{d} \frac{v_{(k)}}{\|\boldsymbol{v}\|_1} \cdot \left( \frac{d - k + 0.5}{d} \right),
\label{eq:gini}
\end{equation}
where \( \|\boldsymbol{v}\|_1 = \sum_{i=1}^d v_i \) is the \( \ell_1 \)-norm.

To measure the sparseness of an attribution vector \( \boldsymbol{\phi}^{(k)} \), we apply the Gini Index to the vector of its absolute values:

\begin{equation}
\text{Sparseness}\left( \boldsymbol{\phi}^{(k)} \right) = G\left( \left| \boldsymbol{\phi}^{(k)} \right| \right),
\end{equation}
where \( \left| \boldsymbol{\phi}^{(k)} \right| = \left( |\phi^{(k)}_1|, |\phi^{(k)}_2|, \ldots, |\phi^{(k)}_d| \right) \). Higher values indicate greater sparseness. A value of 1 implies that the attribution is entirely concentrated on a single feature, while 0 corresponds to equal attribution across all features.

\section{Results}

\subsection{Structural patterns identified using feature engineering based on PCA}\label{sec:pahse1results}
\begin{figure}
\centering
\includegraphics[width=1.\textwidth]{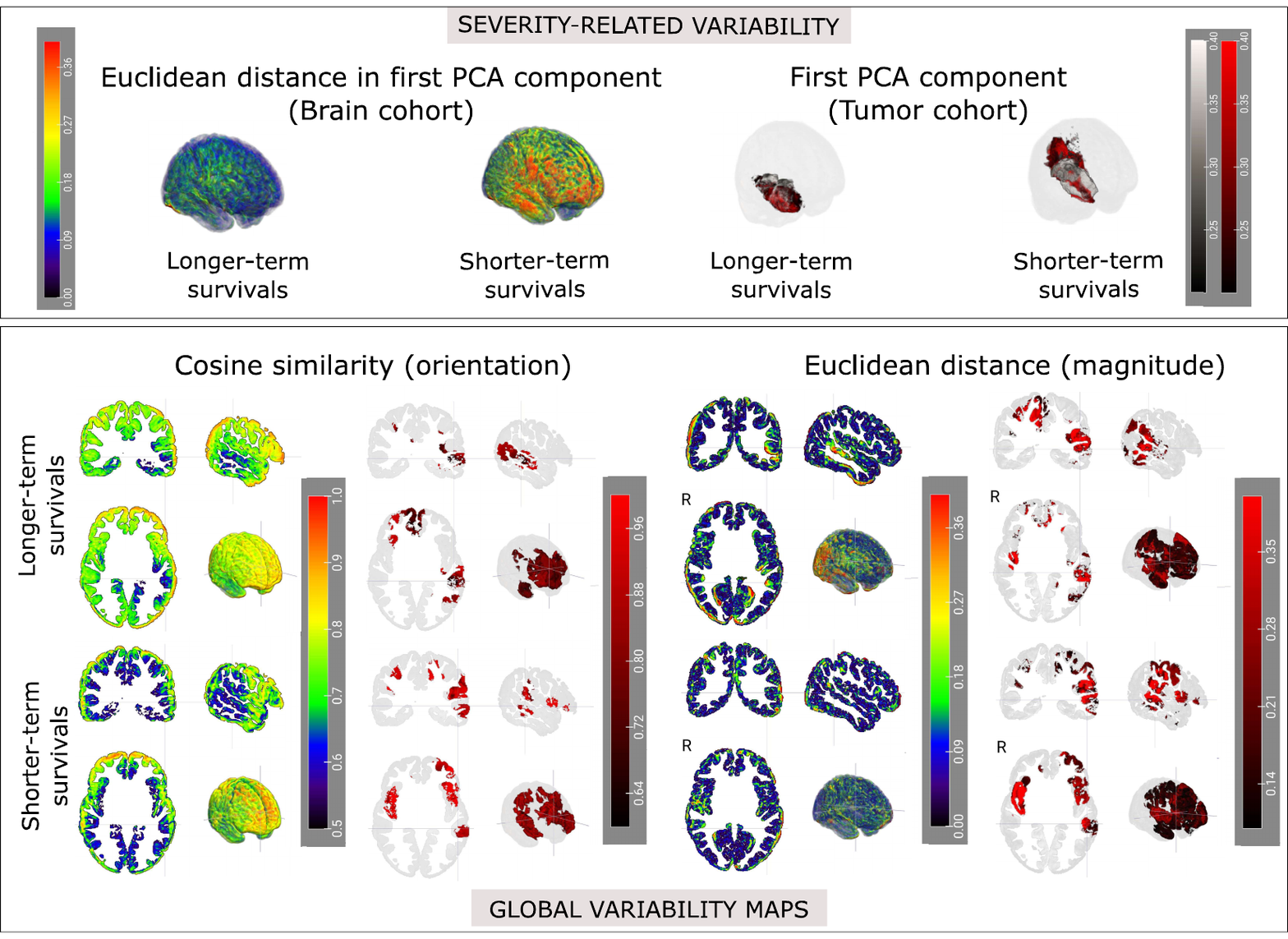}
\caption{Variability across severity conditions. Top: First PCA components analysis showing atlas-based Euclidean distances in the new space between pre- and post-surgery subgroups in the brain cohort. The first PCA distribution is presented for the tumor cohort before surgery (grayscale) and after surgery (red). Bottom: Global variability in PCA components, displaying magnitude and orientation results for the comparison between pre- and post-surgery groups for both brain and tumor cohorts.} \label{fig4}
\end{figure}

Once the PCA components (brain: 8, tumor: 12) were computed across groups, structural variability was quantified to explore spatial differences in tumor and brain patterns. The localization of the first PCA component in the tumor cohorts within the cerebral space is illustrated in the top right of Fig.\ref{fig4}, revealing group-specific spatial distributions. To evaluate brain-wide structural changes, voxel-wise Euclidean distances were computed on the first PCA component, producing variability maps across groups (Fig.\ref{fig4}, top left). The shorter-term survival group showed greater distances between pre- and post-surgery scans, suggesting more pronounced structural alterations. Moreover, this group exhibited higher spatial variability in the tumor PCA component, both before (grayscale) and after surgery (red), suggesting increased heterogeneity in tumor location and size.

To quantify global structural variability, we computed voxel-wise Euclidean distances across PCA components, generating variability maps that highlight key differences between groups. This approach allowed us to capture the overall magnitude of structural differences at each voxel, revealing patterns of brain alterations associated with disease progression. To ensure a robust characterization, we evaluated both the magnitude and orientation of variations in PCA space, comparing pre- and post-surgery subgroups to assess changes relative to disease severity. These maps are displayed in Fig.~\ref{fig4} (\textit{Global Variability Maps} section) and offered a global depiction of structural variability across the brain, highlighting areas where voxel-wise differences between pre- and post-surgery scans were most pronounced in each survival group. 

To identify the most relevant brain regions, we used atlas-based segmentation and applied both intensity and volume criteria. For the Euclidean distance maps, a brain region was considered significant if it met two conditions: it contained at least one voxel above the 95th percentile (indicating a strong local effect), and at least $50\%$ of its voxels exceeded the 80th percentile (reflecting a substantial spatial extent). For the cosine similarity analysis, we focused on regions with the lowest similarity values, as they reflect the greatest divergence in directionality of the PCA patterns. Specifically, we selected regions where the lowest voxel values fell below the 5th percentile, and applied a volume threshold of the 20th percentile to ensure spatial relevance.

Columns \textit{Euclidean maps} and \textit{Cosine maps} of Table~\ref{tab1} summarize the key brain regions identified through PCA-based feature engineering. These regions exhibit the greatest dissimilarity between pre- and post-surgery states in both the longer-term and shorter-term survival groups (\textit{Brain regions} rows), as well as the largest changes observed within the tumour masks before and after surgery (\textit{Surgical regions} row), reflecting differences between tumour locations and the surgical removal area.
\begin{table}
\centering
\resizebox{\textwidth}{!}{%
\begin{tabular}{cc|c|c|c|c}
\hline
 &
   &
  \textbf{PCA Euclidean maps} &
  \textbf{PCA cosine maps} &
  \textbf{\begin{tabular}[c]{@{}c@{}}First PCA from\\ local explanations\end{tabular}} &
  \textbf{\begin{tabular}[c]{@{}c@{}}Global optimizer \\ explanation\end{tabular}} \\ \hline
\multicolumn{1}{c|}{\multirow{2}{*}{\textbf{\begin{tabular}[c]{@{}c@{}}Longer-term \\ survivals\end{tabular}}}} &
  \textbf{\begin{tabular}[c]{@{}c@{}}Surgery \\ regions\end{tabular}} &
  DLP,  EA, IFO, PLMC &
  AA, ACMP, OPF &
  N/A &
  N/A \\ \cline{2-6} 
\multicolumn{1}{c|}{} &
  \textbf{\begin{tabular}[c]{@{}c@{}}Brain \\ regions\end{tabular}} &
  ACMP, EA &
  \begin{tabular}[c]{@{}c@{}}AA, DSV, EA, IFO, \\ LT, MT, PC, VSV\end{tabular} &
  \begin{tabular}[c]{@{}c@{}}AA, IF, LT, MT, \\ MTV, OPF, Premotor\end{tabular} &
  \begin{tabular}[c]{@{}c@{}}AA,ACMP, DLP, IFO,MT,\\  OPF, PC,PLMC,PO,SP,VSV\end{tabular} \\ \hline
\multicolumn{1}{c|}{\multirow{2}{*}{\textbf{\begin{tabular}[c]{@{}c@{}}Shorter-term \\ survivals\end{tabular}}}} &
  \textbf{\begin{tabular}[c]{@{}c@{}}Surgery \\ regions\end{tabular}} &
  ACMP, EA, IFO, PO &
  DLP, OPF &
  N/A &
  N/A \\ \cline{2-6} 
\multicolumn{1}{c|}{} &
  \textbf{\begin{tabular}[c]{@{}c@{}}Brain \\ regions\end{tabular}} &
  EA, IFO, OPF, PC &
  \begin{tabular}[c]{@{}c@{}}EA, IFO, MT,\\  MTV, PC, PO, VSV\end{tabular} &
  \begin{tabular}[c]{@{}c@{}}DSV, IF, IFO, \\ MTV, VSV\end{tabular} &
  \begin{tabular}[c]{@{}c@{}}AA,EA,MT,OPF,\\ PC,PO,VSV\end{tabular} \\ \hline
\end{tabular}%
}
\vspace{1mm} 
\noindent\parbox{0.85\textwidth}{%
\fontsize{8pt}{10pt}\selectfont
AA: Auditory Association, ACMP: Anterior Cingulate and Medial Prefrontal,  DLP: Dorsolateral Prefrontal, DSV: Dorsal Stream Visual, EA: Early Auditory, IF: Inferior Frontal, IFO: Insular and Frontal Opercular, LT: Lateral Temporal, MT: Medial Temporal, MTV: MT+ Complex and Neighboring Visual Areas, OPF: Orbital and Polar Frontal, PC: Posterior Cingulate, PLMC: Paracentral Lobular and Mid Cingulate, PO: Posterior Opercular, SP: Superior Paretal, VSV: Ventral Stream Visual.
}
\caption{Key brain regions with significant 3D volume differences pre- vs. post-surgery and surgical regions highlighting dissimilarities between tumor volumes and surgical removal areas across survival groups.}\label{tab1}
\end{table} 

\subsection{Ablation Study of Unsupervised Pretraining and Fine-Tuning Strategies}
We conducted an ablation study comparing two different cohort training strategies (see \ref{sec:xai_1}). Based on Table \ref{tab:metrics}, the best validation results were achieved using the second cohort strategy, particularly with the Addenbrooke’s Hospital and BraTS2025 datasets. Specifically, the Swin-Unet model achieved the lowest error values across both strategies, with the best performance in the second strategy; an RMSE of 0.008 compared to 0.010, MSM of 0.001 in both cases, and MAE of 0.005 compared to 0.009. These results highlight the superiority of the strategy involving both the Addenbrooke’s Hospital and BraTS2025 datasets over the strategy using only the Addenbrooke’s Hospital dataset.
\begin{table}[]
\centering
\resizebox{\textwidth}{!}{%
\begin{tabular}{c|cc|cc}
\hline
\multirow{2}{*}{} & \multicolumn{2}{c|}{Addenbrooke’s Hospital}                        & \multicolumn{2}{c}{Addenbrooke’s Hospital and BraTS2025} \\ \cline{2-5} 
            & \multicolumn{1}{c|}{Swin-Unet}             & Autoencoder           & \multicolumn{1}{c|}{Swin-Unet} & Autoencoder \\ \hline
Training loss     & \multicolumn{1}{c|}{0.020 \textpm{} 0.004} & 0.040 \textpm{} 0.003 & \multicolumn{1}{c|}{0.003}            & 0.017            \\ \hline
Validation loss   & \multicolumn{1}{c|}{0.040 \textpm{} 0.050} & 0.060 \textpm{} 0.030 & \multicolumn{1}{c|}{0.004}            & 0.018            \\ \hline
RMSE metric & \multicolumn{1}{c|}{0.010 \textpm{} 0.005} & 0.020 \textpm{} 0.005 & \multicolumn{1}{c|}{0.008}     & 0.020       \\ \hline
MSM metric  & \multicolumn{1}{c|}{0.001 \textpm{} 0.002} & 0.001 \textpm{} 0.001 & \multicolumn{1}{c|}{0.001}     & 0.001       \\ \hline
MAE metric  & \multicolumn{1}{c|}{0.009 \textpm{} 0.007} & 0.019 \textpm{} 0.006 & \multicolumn{1}{c|}{0.005}     & 0.017       \\ \hline
\end{tabular}%
}
\vspace{1mm}
\noindent\parbox{0.9\textwidth}{%
\fontsize{8pt}{10pt}\selectfont
RMSE: Root Mean Squared Error, MSM: Mean Squared Magnitude, MAE: Mean Absolute Error.
}
\caption{Training and validation metrics from unsupervised learning of structural MRI, 5-fold cross-validation was used in the Addenbrooke’s Hospital case and 60\% - 40\% training validation split in the Addenbrooke’s Hospital and BraTS2025.}
\label{tab:metrics}
\end{table}

\begin{figure}
\centering
\begin{subfigure}[t]{\columnwidth}
    \centering
    \includegraphics[width=\columnwidth]{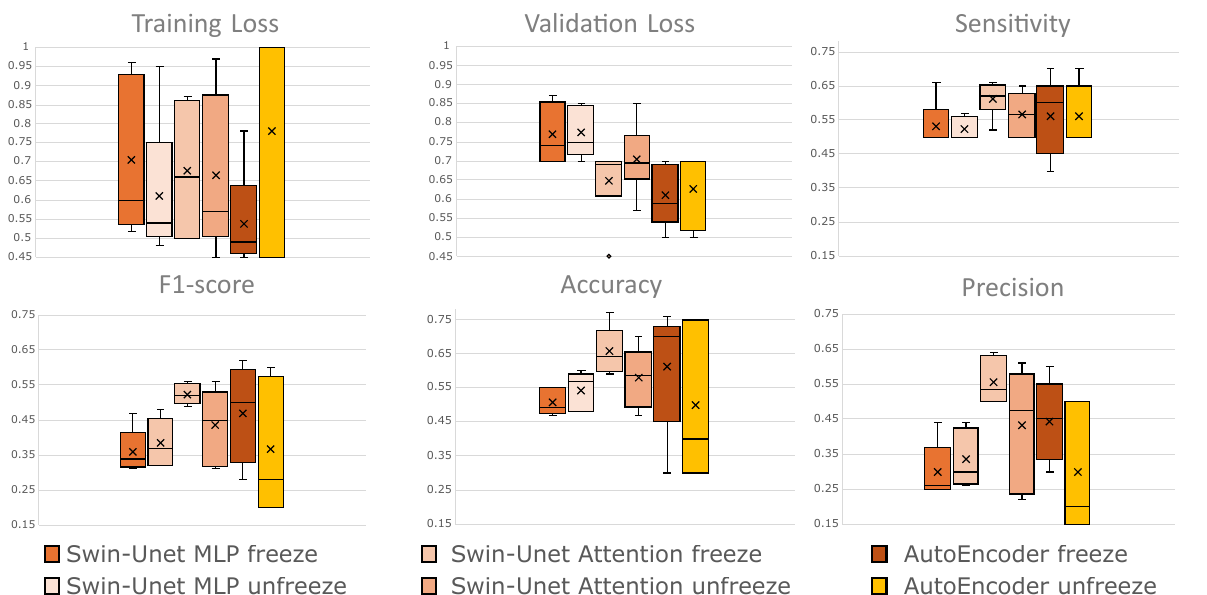}
    \caption{Ablation study evaluating incorporating MLP and attention modules under encoder freeze and unfreeze strategies during fine-tuning. Metrics highlight the impact of architectural choices and training configurations.}
    \label{fig:xai2}
\end{subfigure}
\begin{subfigure}[t]{\columnwidth}
    \centering
\includegraphics[width=\columnwidth]{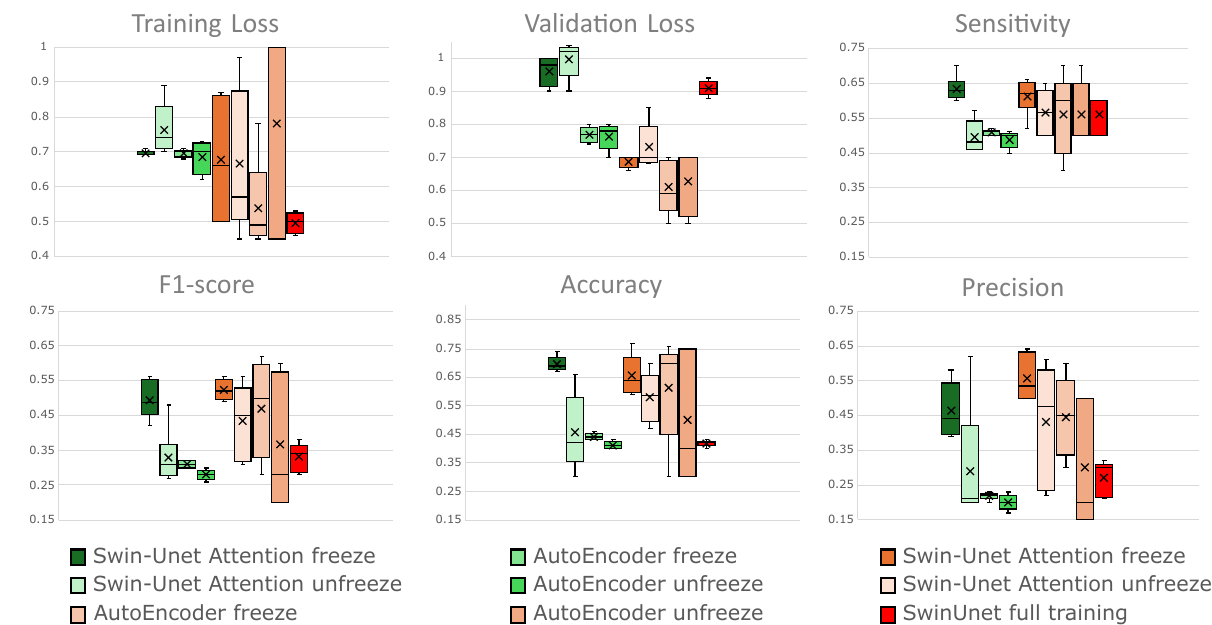}
    \caption{Ablation study under different unsupervised training cohort configurations, including encoder freezing and full training scenarios. Metrics demostrate the impact of training strategies on model performance.}
    \label{fig:classif}
\end{subfigure}
\caption{Examples of training and validation results in the ablation study of Swin-Unet and AutoEncoder variants (a) during fine-tuning with frozen and unfrozen encoder settings; (b) under different unsupervised training configurations.}
\label{fig32}
\end{figure}

The performance outcomes for fine-tuning in the survival binary classification task using sMRI data (see \ref{sec:binary}) are illustrated in Fig.~\ref{fig32}. We conducted an ablation study across three encoder-decoder configurations: Swin-Unet with an MLP output layer, Swin-Unet with an attention-based output layer, and a baseline AutoEncoder. Each model was evaluated under two fine-tuning strategies: (i) freezing the pre-trained encoder, and (ii) unfreezing the encoder during downstream training. The variability reported reflects results obtained via 5-fold CV.
As shown in Fig.~\ref{fig32}a, frozen encoders exhibited higher variability across most metrics compared to their unfrozen counterparts. Surprisingly, the frozen configurations also achieved higher average performance. Among all models, the Swin-Unet with an attention-based output layer and frozen encoder achieved the best overall results, with an average F1-score of 0.52, accuracy of 0.67, sensitivity of 0.64, and precision of 0.55. Its maximum values across folds reached an F1-score of 0.56, accuracy of 0.77, sensitivity of 0.66, and precision of 0.65.
Although the AutoEncoder architecture achieved slightly higher maximum values in F1-score and sensitivity, it exhibited considerably higher CV variability across all metrics and consistently lower precision (below 0.57), indicating a higher false-positive rate compared to the Swin-Unet with the attention-based output layer. These findings highlight a trade-off between performance stability and sensitivity, and suggest that attention-based decoding in transformer-style architectures offers a more reliable and interpretable solution for domain-specific fine-tuning in neuroimaging applications.

Lastly, Fig.~\ref{fig32}b provides further evidence from the ablation study on different unsupervised training cohorts in the fine-tuning task, confirming the superiority of the strategy that leverages both the Addenbrooke’s Hospital and BraTS2025 datasets compared to the approach that relies solely on the Addenbrooke’s Hospital dataset. Training from scratch on the Addenbrooke’s Hospital dataset without any fine-tuning resulted in substantially poorer performance compared to either of the two unsupervised training cohorts strategies.

\subsection{Interpretable Deep Learning for Survival Classification}
As the Swin-Unet model with an attention-based output layer and a frozen, unsupervised pre-trained encoder trained on both the Addenbrooke’s Hospital and BraTS2025 datasets outperformed all other configurations, we applied the explanation framework exclusively to this model.
\subsubsection{Metrics and interpretations of XAI models} \label{xai_results}

\begin{table}[]
\centering
\resizebox{\textwidth}{!}{%
\begin{tabular}{c|c|c|c|c|c}
\hline
Method             & RMSE             & MAE              & MSM              & Sparseness       & Faithfulness     \\ \hline
Global optimizer (proposed) & \textbf{0.964 \textpm{} 0.12} & \textbf{0.610 \textpm{} 0.11} & \textbf{0.967 \textpm{} 0.22} & 0.537 \textpm{} 0.31 & \textbf{0.913 \textpm{} 0.04} \\ \hline
Gradient SHAP  & 1.066 \textpm{} 0.20 & 0.665 \textpm{} 0.22 & 1.160 \textpm{} 0.47 & 0.441 \textpm{} 0.01 & 0.370 \textpm{} 0.38 \\ \hline
Guided Backpropagation  & 1.061 \textpm{} 0.21 & 0.678 \textpm{} 0.26 & 1.175 \textpm{}0.46 & 0.427 \textpm{} 0.01 & 0.380 \textpm{} 0.17 \\ \hline
Guided GradCam & 1.067 \textpm{} 0.20 & 0.643 \textpm{} 0.19 & 1.166 \textpm{} 0.47 & \textbf{0.611 \textpm{} 0.05} & 0.362 \textpm{} 0.31 \\ \hline
Input X Gradient & 1.095 \textpm{} 0.12 & 0.674 \textpm{} 0.26 & 1.189 \textpm{} 0.35 & 0.10 \textpm{} 0.02 & 0.273 \textpm{} 0.19 \\ \hline
Integrated Gradient & 1.095 \textpm{} 0.12 &  0.681 \textpm{} 0.25 & 1.189 \textpm{} 0.35 & 0.445 \textpm{} 0.01 & 0.386 \textpm{} 0.26 \\ \hline
Kernel SHAP & 1.095 \textpm{} 0.15 & 0.690 \textpm{} 0.23 & 1.189 \textpm{} 0.37 & 0.444 \textpm{} 0.01 & 0.35 \textpm{} 0.16 \\ \hline
\end{tabular}%
}
\vspace{1mm}
\noindent\parbox{0.9\textwidth}{%
\fontsize{8pt}{10pt}\selectfont
RMSE: Root Mean Squared Error, MSM: Mean Squared Magnitude, MAE: Mean Absolute Error.
}
\caption{Training and validation metrics from Global explanations of structural MRI.}
\label{tab:metrics2}
\end{table}

The proposed global explanation optimizer outperformed both the baseline explanation methods used during its training and testing; namely, Gradient SHAP, Guided Backpropagation, and Guided Grad-CAM, as well as established explanation techniques not involved in its training process, including Input × Gradient \cite{1}, Integrated Gradients\cite{1}, and Kernel SHAP \cite{gradshap}. In terms of faithfulness, the optimizer achieved a score of 0.913 (see Table~\ref{tab:metrics2}). It also had the lowest average RMSE (0.964), MAE (0.610), and MSM (0.967). Standard deviation was assessed across four global explanations: pre- and post-surgery as well as shorter-term and longer-term survivals. While Guided GradCam showed the highest sparseness (0.612), its faithfulness was below 0.362. The optimized method had the highest reliability aligning closely with the first PCA component of sMRI images and preserving key PCA-derived features. Fig.\ref{fig:xai} illustrates results for the post-surgery longer-term survivals and pre-surgery shorter-term survivals using different $l_1$, $l_2$, and $l_3$ parameter combinations from \eqref{3}.

\subsubsection{Patterns identified in XAI explanations}

\begin{figure*}
\centering
\includegraphics[width=0.95\textwidth]{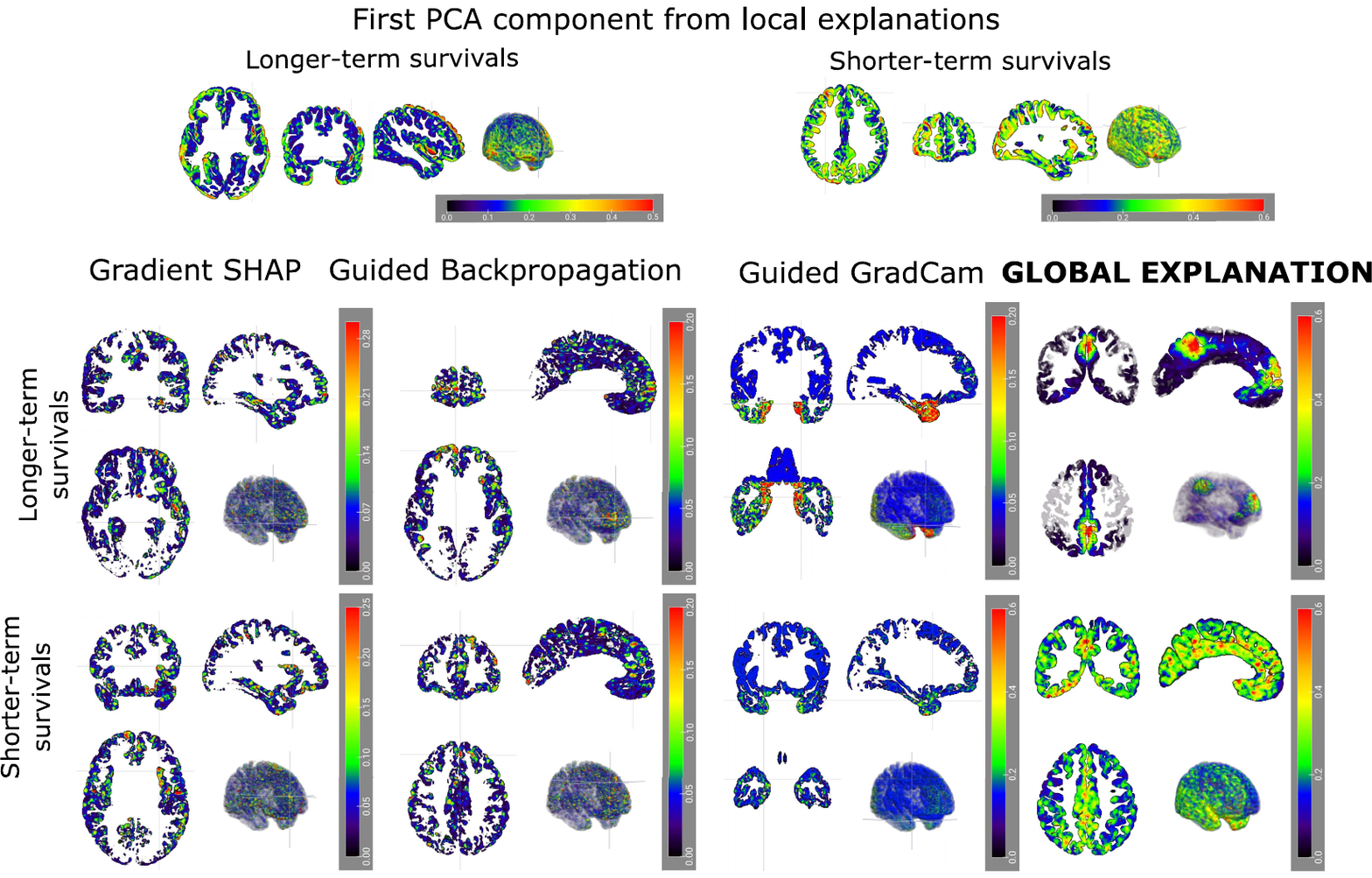}
\caption{Explainability analysis results. For each severity condition (shorter-term and longer-term survivals), multi-planar slices  of the representations obtained are displayed. Top:  the first PCA component. Bottom: from left to right the outcomes of applying gradient SHAP, guided backpropagation, guided Grad-CAM, and the final result associated with the global explanation optimizer. } \label{fig5}
\end{figure*} 

Fig.~\ref{fig5} displays the Euclidean distances between pre- and post-surgery scans for both the longer-term and shorter-term survival cohorts. These distances are shown for the first PCA component derived from local explanations, the local explanations themselves (\textit{Gradient SHAP}, \textit{Guided Backpropagation}, and \textit{Guided GradCAM}), as well as for the global explanation (\textit{Global explanation}). The global explanation optimizer outperforms the other methods in terms of sparsity and faithfulness, offering better insights into the global patterns. Thus, we focus primarily on discussing the results from the first PCA component obtained from local explanations and the global explanation maps in Fig.~\ref{fig5}. By comparing with the atlas using the same thresholding criterion as applied in the PCA Euclidean distance maps, at least 50\% of voxels exceeding the 80th percentile and at least one voxel above the 95th percentile, the significant regions are summarised in Table~\ref{tab1} (columns \textit{First PCA from local explanations} and  \textit{Global optimizer explanation}). 

A suggested guidance based on Table~\ref{tab1} follows the pattern below: Feature engineering (PCA-based Euclidean and cosine maps) revealed that the \textit{surgery regions}, i.e. those showing the largest pre- vs. post-surgery changes within tumour masks PCA-space, partially overlap with (and may help explain) the post-surgery alterations observed in \textit{brain regions}. A key example is the Early Auditory (EA) cortex, which consistently appeared across both survival groups and map types in both the surgical and brain-level results, suggesting it is a core region affected by tumour resection and a hub of post-operative reorganisation \cite{kokkonen2005effect}. Similarly, Insular and Frontal Opercular (IFO) areas and the Orbital Polar Frontal (OPF) cortex were commonly involved, indicating that disruption to sensory and frontal integration areas may play a central role in shaping global connectivity changes \cite{fang2016influence}. In longer-term survivors, surgical effects were more confined to frontal and midline structures (e.g. Anterior Cingulate and Medial Prefrontal, ACMP), with downstream changes in executive and motor regions, possibly engaging compensatory networks such as the frontoparietal control system. In contrast, shorter-term survivors showed surgical involvement in posterior and multimodal sensory areas (e.g. Posterior Opercular, PO, or Ventral Stream Visual, VSV), paralleled by more diffuse alterations in visual and perceptual cortices, which may reflect greater network fragility or reduced plasticity.

The final two columns of Table~\ref{tab1}, representing local and global explanation methods, further highlight the regions most relevant for binary survival classification. Among longer-term survivors, local explanations emphasized frontal and temporal areas such as the Inferior Frontal (IF), Lateral Temporal (LT), and Medial Temporal (MT) regions \cite{niki2020primary}. These regions support language, memory, executive functions, and motor planning, consistent with preserved or adaptable networks facilitating recovery. Global explanations in the same group additionally highlighted integrative hubs such as ACMP, Dorsolateral Prefrontal Cortex (DLP), MT, and OPF, which are associated with emotional regulation, high-order cognition, and multisensory integration \cite{hornak2004reward}. In shorter-term survivors, the first PCA component from local explanations also included frontal and MT regions, but greater emphasis was placed on posterior sensory and association cortices such as the Dorsal Stream Visual (DSV) and VSV, suggesting stronger disruption of visual and interoceptive systems that may be less amenable to functional compensation.
\begin{figure*}
\centering
\includegraphics[width=0.7\textwidth]{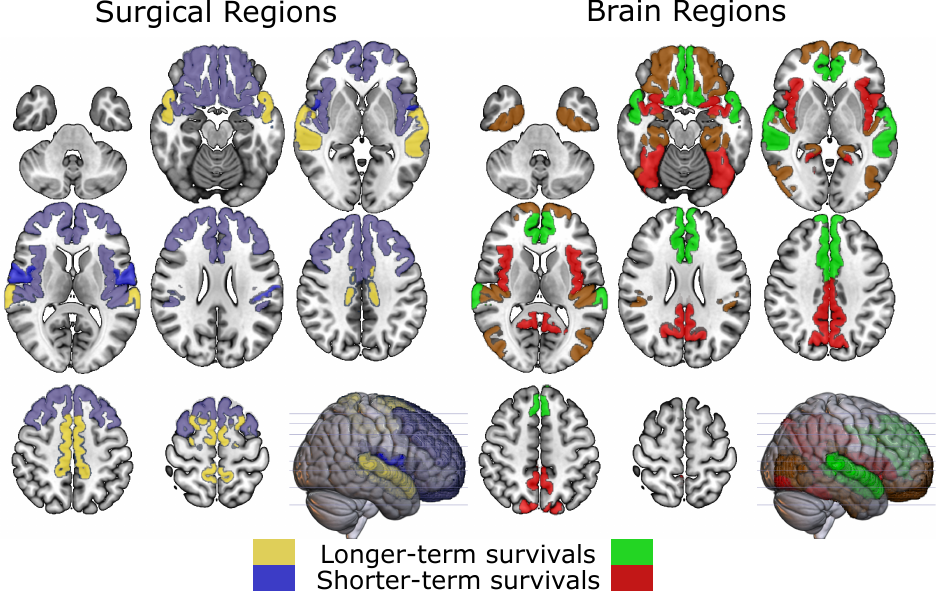}
\caption{Surgical and brain regions with the most significant changes according to the combined framework for longer-term and shorter-term survival groups. Surgical regions include all areas identified across frameworks, while brain regions are limited to those consistently detected by at least two different frameworks within the same survival group. } \label{fig_high}
\end{figure*}
Taken together, the patterns identified across PCA-derived feature maps and model explanation methods suggest that focal surgical changes, particularly in regions such as EA, IFO, and OPF—are linked to broader alterations in structurally and functionally connected brain areas. Longer-term survivors showed more consistent engagement of fronto-cingulate and temporal regions (e.g., DLP, ACMP, PLMC), associated with executive and cognitive control functions, while shorter-term survivors exhibited widespread posterior sensory and visual involvement (e.g., PO, DSV, VSV), consistent with less focal and less compensable network disruption \cite{nakajima2019glioma}. Fig.~\ref{fig_high} illustrates the regions most consistently differentiated between survival groups. The recurrence of hubs such as EA and OPF across multiple analytical methods underscores their centrality in post-operative adaptation and suggests that surgical impact on specific cortical hubs may shape the extent of functional reorganisation, thereby influencing clinical outcomes.

\section{Discussion}

In this study, we studied how structural brain reorganization after glioma surgery relates to patient survival, leveraging a uniquely curated and rare dataset of paired pre- and post-surgical MRI scans. By integrating XAI with latent-space feature engineering (PCA), we identified survival-related neuroimaging biomarkers and generated cohort-level explanations that improved both interpretability and reliability compared to existing state-of-the-art methods.

Our findings demonstrate that regions undergoing the greatest surgical changes often overlapped with broader post-surgical alterations in brain networks, underscoring the interplay between local resection effects and global connectivity reorganization. A consistent involvement of the Early Auditory (EA) cortex across both survival groups suggests that it may represent a central hub of post-operative plasticity as well as a point of vulnerability \cite{rauschecker1999auditory}. Other regions, including the Insular, Frontal Opercular, and OPF cortices, were repeatedly implicated, highlighting the key role of sensory–frontal integration areas in shaping recovery trajectories \cite{duffau2006insular,mitolo2022neuroplasticity}.
Clear survival-related distinctions emerged. Longer-term survivors exhibited more localized surgical effects in frontal and midline structures, with downstream engagement of executive and motor networks, possibly reflecting the recruitment of compensatory systems such as the frontoparietal control network. In contrast, shorter-term survivors displayed greater involvement of posterior and multimodal sensory areas, together with diffuse alterations in visual and perceptual cortices. These contrasting profiles suggest differences in network resilience and neuroplasticity, pointing to potential imaging markers for surgical planning, risk stratification, and post-operative rehabilitation strategies \cite{cargnelutti2020we}. Overlap patterns between explanatory methods further support these distinctions. For the longer-term survival cohort, both the first PCA component and the global optimizer highlighted Orbital and Polar Frontal regions, whereas in the shorter-term survival cohort, overlap was observed in the Auditory Association, MT, and OPF regions. Importantly, the superiority of the global explanation optimizer is evident in its ability to identify additional survival-related hubs, such as the Posterior Cingulate in shorter-term survivors and the ACMP regions in longer-term survivors, which were not captured by the first PCA component alone (see Section~\ref{sec:pahse1results}, Fig.~\ref{fig5}). These findings illustrate how optimized global explanations can provide more consistent and clinically meaningful insights compared to conventional approaches.

The main limitation of this work lies in the restricted availability of paired pre- and post-surgical structural MRI data. Such longitudinal imaging remains extremely scarce in clinical practice due to the clinical, logistical, and ethical challenges of acquisition. Although large public datasets are increasingly accessible, they typically lack longitudinal follow-up or rely on synthetic or heavily preprocessed data, which may not adequately reflect clinical variability. By contrast, our dataset consists of real-world clinical cases collected under routine care, capturing the heterogeneity, imaging artefacts, and surgical effects that are often absent in curated repositories. This rarity constitutes a unique strength of the study, enabling us to more accurately model the structural consequences of surgery and better capture individual variability in post-operative brain reorganization. To mitigate the limitations of sample size, we are actively expanding the collection of longitudinal cases to strengthen statistical power and reproducibility. Future work will also extend the framework to additional neuroimaging modalities, including functional and diffusion MRI and diffusion MRI. These complementary modalities can enhance predictive performance, enrich interpretability, and provide a more comprehensive view of both structural and functional brain dynamics. Integrating multimodal imaging perspectives will ultimately advance our understanding of surgical impact, recovery mechanisms, and disease progression, and further reinforce the translational potential of XAI-driven neuroimaging in precision neuro-oncology.

\section{Conclusions}

Our proposed framework integrates XAI with neuroimaging-based feature engineering to predict survival in brain tumor patients, offering guidance for surgical decision-making to achieve the critical onco-functional balance. A unique strength of this work lies in the use of a rare, clinically collected dataset comprising paired pre- and post-surgical MRI scans—data that are exceptionally scarce due to the clinical complexities, operational demands, and ethical considerations of acquiring longitudinal datasets. Using this unique dataset, we demonstrate how dissimilarities between tumor volumes and surgical resection areas correlate with their structural impact on the brain post-operatively.
By extracting global explanations from deep learning models for predicting short- and long-term survival, the framework functions as a clinically relevant predictive guideline. Our results highlight the consistent involvement of sensory and cognitive regions, with greater disruptions observed in shorter-term survivors, underscoring the importance of preserving networks critical for cognition and perception. Methodologically, the proposed global explanation optimizer improves both faithfulness and interpretability compared to alternative global XAI methods, while reducing inter-method variability that often undermines the trustworthiness of explainable AI.
Overall, this work not only establishes a novel XAI-based framework for survival assessment but also demonstrates the scientific and clinical value of rare pre- and post-surgical datasets in uncovering survival-related variability, ultimately advancing precision medicine in neuro-oncology.

\section*{Acknowledgment}
This project has received funding from the European Union's Horizon Europe research and innovation program under grant agreement No 101147319 (EBRAINS 2.0). It was also supported in part by the PID2022-137629OA-I00 and PID2022-137451OB-I00 projects, funded by the MICIU/\allowbreak AEI/\allowbreak 10.13039/\allowbreak  and by ``ERDF/EU''. This study was funded from the National Institute for Health and Care Research (NIHR), Career Development Fellowship (CDF-2018-11-ST2-003 to SJP.). This publication presents independent research funded by the National Institute for Health and Care Research (NIHR). All research at the Department of Psychiatry in the University of Cambridge is supported by the NIHR Cambridge Biomedical Research Centre (NIHR203312) and the NIHR Applied Research Collaboration East of England. The views expressed are those of the author(s) and not necessarily those of the NHS, the NIHR or the Department of Health and Social Care. C.J.M is supported by grant JDC2023-051807-I funded by MICIU/AEI/10.13039/501100011033 and by ESF+.

\bibliographystyle{unsrt}  
\bibliography{glioma_bib}  

\end{document}